\documentstyle[preprint,aps,prl,12pt,epsf]{revtex}

\begin{document}
\draft

\title{Non-linear mean field dynamics in the nuclear spinodal zone}

\author{ G.F. Burgio, M. Baldo, and A. Rapisarda}

\address{  Universit\'a di Catania and INFN Sezione di Catania}
\address{  Corso Italia 57, I-95129 Catania, Italy }
\date{June 4th 1993 - Catania University preprint no.93/20}
\maketitle

\begin{abstract}
We demonstrate, by numerical simulations, that the dynamics of nuclear
matter mean field inside the spinodal region is chaotic.
Spontaneous symmetry-breaking - no explicit fluctuating term is
considered - occurs leading to wild unpredictable density fluctuations.
A proper recipe to calculate an average Lyapunov exponent in this
multidimensional phase space is introduced. The latter is
calculated for different values of the density in order to characterize
in a quantitative way the chaotic and regular regions.
It is argued that the mean field chaoticity can be
the main mechanism
of the nuclear multifragmentation occurring in the intermediate
energy reactions.

\end{abstract}



Heavy ion collisions in the intermediate energy regime
between 20 and
100 MeV/nucleon can result in multifragment emissions, where a number
of intermediate mass fragments (IMF : Z $\geq$ 2) is observed. This
disassembly of the nuclear system, which can include a substantial
fraction of the whole colliding system, falls under the generic name
of  "nuclear multifragmentation", and is presently under intense study
in different laboratories. The main mechanism which is responsible
of this phenomenon has not yet been determined. Both statistical models
and dynamical simulations \cite{gro93} have been used to
describe the cluster formation in the multifragmentation regime.
The role of mechanical instabilities, either of the Raleigh type
\cite{mor92}, or associated with ring
formation \cite{bau92}, has been also stressed recently. These
calculations provide suggestions on the possible dominant path
followed by the nuclear system along the multifragmentation process.
The onset of instability in the spinodal zone of the nuclear matter
Equation of State (EOS) has been studied in ref. \cite{hei93},
by analyzing the corresponding linear response for imaginary frequency
as a function of the wave number and of the density.
\par
In this letter we establish, by numerical simulations,
that the dynamical evolution of nuclear
matter mean field in the spinodal region is chaotic. We estimate
the corresponding Lyapunov exponent which characterizes the rate of
divergence between the trajectories in density space. This result
can have far reaching
consequences for the whole picture of the multifragmentation process.
In particular, a) the relevance of the mean trajectory,
which is an essential ingredient in most of the simulations with kinetic
equations, appears
questionable if the spinodal region is reached; b) the onset of
chaoticity
implies, as we explicitly verify numerically, strong non--linearity of
the evolution, and therefore strong coupling between different modes;
c) the occurrence of chaotic motion suggests that thermodynamical
quantities, like entropy, are not necessarily appropriate for
characterizing the system dynamics, but they
should be substituted by others, based on chaotic evolution concepts,
like the Kolmogorov entropy \cite{ben76}. Other more quantitative
consequences will be discussed in the following.
The numerical analysis is performed by solving the Vlasov equation
\begin{equation}
\label {1}
{\partial f \over \partial t} + \{H[f], f\} =0
\end{equation}
\noindent
on a two-dimensional lattice, according to the method of ref.\cite{bur92}.
In eq. (1) $H[f]$ is the self-consistent effective one-body
Hamiltonian. In this approach the one-body density distribution is
represented  on a grid at variance with the more commonly used test
particle methods \cite{gre87,ber88}. The former has been preferred
since it
allows a better control of the numerical accuracy and conservation laws.
The single particle phase space is divided into cells and nuclear matter
is confined in a large torus with periodic boundary conditions.
The side lengths of the torus are equal to $L_x=51~fm$ and  $L_y=15~fm$.
In order to achieve sufficient accuracy,
very small cells having $\delta x = {1\over3}$ fm and $\delta p=40$ MeV/c
are employed. The time integration step was chosen equal to 0.5 fm/c.
We adopt a simplified Skyrme interaction for the effective one-body field
which is averaged over the y-direction \cite{bur92}.
We employ a Fermi momentum at saturation of $P_F=260$ MeV/c, which gives a
density in two dimensions equal to $\rho_0=0.55 ~fm^{-2}$.
The mean field evolution is treated by means of a standard matrix
technique \cite{bur92}.
The main indication of chaotic behaviour of a dynamical system is the
extreme sensitivity of its evolution to the initial conditions. This is
illustrated in figs. 1a) and 1b).  At the initial time nuclear matter
presents a density wave, with an amplitude $\Delta\rho_{1}$ =
10$^{-2}\rho_0$, and with a wave number $k = 2\pi (5/L_x)$.
The local momentum distribution is assumed to be the one
of a Fermi gas at a temperature of 3 MeV. This corresponds to
a very small zero sound oscillation. The initial conditions are
varied in a very simple way, namely by changing the density $\rho$
by an amount $\Delta\rho = 10^{-4}\rho_0$. No appreciable change of
the results are obtained with different prescriptions of modifying
the initial conditions, as discussed later. We checked that in all
the calculations here presented the total energy was conserved up
to 1$\%$ and that the particle number remains constant during the
whole evolution.
For a density equal to the saturation density $\rho_0$,
fig.1 (a), the density profiles of nuclear matter corresponding
to different initial conditions keep close to each other for the whole
considered time interval. This is typical of the "regular" region
of a dynamical system, and indicates the stability of the dynamics
with respect to small perturbations. On the contrary, if the
density $\rho$ is chosen
well inside the spinodal region, in this case $\rho = {1 \over 2}\rho_0$,
the density profiles, after a short time, differ
completely between each other, even if initially they are close within
a part over $10^4$.
This is the typical behaviour of a dynamical system in a chaotic
region, where very small initial
perturbations are rapidly amplified.
The relevant feature of fig.1 (b) is that the two density profiles do not
differ simply in their amplitudes, as it
would occur if the linear response regime would be valid, but mainly
in their shape, which can occur only in a regime of strong
non--linearity. We have checked that the results
are stable with respect to the increase of the precision in the
numerical integration. In fact, a decrease of the time integration step
does not affect the results, and therefore the divergence in shape
of the trajectories is an intrinsic
dynamical property of the nuclear mean field in this region.
The chaotic dynamics  is intimately related with the strong
non--linearity of the evolution, which implies a strong
coupling among the different modes. This can be checked by Fourier
transforming the density profiles at different times. We found
that the different wave numbers increase their mixing more and more
as the evolution proceeds. At the latest stage
a wide range of wave numbers result to be excited,
despite the fact that only one wave number $k$ is initially populated.
This point is illustrated in fig.2 for a single trajectory
corresponding to the full curve of fig.1 (b).
The final wave number spectral density has therefore little
resemblance with the initial one, and this feature was found to be
independent from the particular initial $k$. In a linear regime,
on the contrary, the system would be equivalent to a set of
independent harmonic oscillators,
one for each wave number $k$, with an imaginary frequency,
corresponding to the unstable region. Such a system is integrable, and
the corresponding flow in phase space is regular.
It is important to note that
the mixing takes place in an interval of time of about 20 $fm/c$,
which is short with
respect to a typical collision time or expansion time, characteristic
of heavy ion collisions at intermediate energy. The calculations
here presented are in two dimensions, but the characteristic times
of chaotic dynamics are expected to be little dependent on
dimensionality.
A behaviour similar to that shown in fig.1 (b) was found for a single
trajectory in the case of ref. \cite{bur92} where a dissipative term
of Langevin type has been added to the Vlasov equation. In principle
a sensitive dependence on the initial conditions should be present
also in that case. Calculations in that direction are in progress.
To perform a more quantitative analysis, we follow the standard procedure
in the theory of dynamical systems of characterizing
the divergence among the phase space trajectories by a set
of Lyapunov exponents \cite{ben76}. Let $d(t)$ be the distance between two
trajectories, along an unstable direction in phase space,
at a given time $t$, and define
\begin{equation}
            \lambda (t)\, =\, {\log (d(t)/d_0) \over t}  ~,
\end{equation}
\noindent
where $d_0 = d(0)$.
Then the corresponding Lyapunov exponent is obtained by the limit

\begin{equation}
          \lambda_{\infty} \, = \, \lim_{d_0\to0}\, \lim_{t\to\infty}\,
 \lambda(t)
\end{equation}
\noindent
provided the limit exists. The very definition of eq. (2) implies that
asymptotically for large time and small enough perturbation
the divergence is essentially exponential, namely that the
equation of motion for the distance between the two trajectories
is a linear one. This by no means implies that each trajectory follows
linear equations of motion.
In numerical applications one has to select a series of small
values of $d_0$, a set of increasing times $t$ and check that the
corresponding values of $\lambda(t)$ stabilize around a definite value
$\lambda_{\infty}$. In the present case
the actual number of degrees of freedom and therefore the number of
Lyapunov exponents is just equal to the number
of cells in which the phase space is divided.
In order to calculate an average value of the Lyapunov exponent,
it is convenient  to choose a distance
which is connected with a global description of the system. We have
therefore chosen as distance between two trajectories the difference in
norm between their density profiles
\begin{equation}
  d(t) \, =\, \sum_i \, | \rho^{(1)}_i(t) - \rho^{(2)}_i(t) |/N_c ~,
\end{equation}
\noindent
where the index $i$ runs over the $N_c$ cells in ordinary space, and
$\rho^{(1)}_i$, $\rho^{(2)}_i$ are the densities in the cell $i$ for
the trajectories $1$ and $2$ respectively. This definition neglects
possible differences which can occur in momentum space between the two
trajectories. The definition of eq. (4) is sufficient for the
present analysis. It should include the contribution of all the unstable
modes, in particular the ones which dominate for large $t$.
In fig.3 (a) it is displayed the time evolution of $\lambda(t)$ for
different values of $\rho/\rho_0$, taking an initial
$d_0 = 10^{-4}\rho$. It was checked that for $d_0$ values of this size,
$\lambda(t)$ stabilizes and is independent of $d_0$. One can distinguish
two well defined regimes. For densities $\rho$ well inside the spinodal
region, the time dependence of $\lambda(t)$ is quite weak, resulting
in the almost straight lines
appearing at the top of fig.3 (a). This behaviour
unambiguously indicates that the Lyapunov exponent in this case is
well defined and that the system evolves in a genuine
chaotic way. As the density increases and approaches the upper limit
of the spinodal region, the general trend starts to change.
In this case
$\lambda(t)$ becomes a decreasing function of $t$
which approaches rapidly very small values,
indicating that the trajectories actually do not diverge between each
other. The general behaviour appearing in fig.3 (a) is typical of dynamical
systems possessing a regular and a chaotic region \cite{ben76}.
The actual value of $\lambda(t)$ can be
considered a measure of the non--linearity and
chaoticity of the system \cite{nota93}. It has to be also stressed that,
when the dynamics
is non--linear, the Lyapunov exponent
$\lambda_{\infty}$ has not to be confused with the rate of increase
of the density fluctuations. In the present case we found
that, for a given trajectory, the
density fluctuations increase with a
characteristic time \cite{bur92} substantially longer than the divergence
characteristic time $1/\lambda_{\infty}$ \cite{bur92}, by a factor
ranging from 2 to 3, according to the density and the initial
conditions. In fig.3 (b) we show  the Lyapunov exponent $\lambda_{\infty}$
calculated at different densities.
For density values close to the spinodal boundary
 $\lambda(t)$ shows some oscillations even for the largest
times considered.
 However in this region the
values are very small  and the general trend of fig.3 (b)
does not depend on this uncertainty. The value of $\lambda_{\infty}$
has a flat maximum around the  density $\rho = 0.35\rho_0$.
The curve reported in fig.3 (b)
is essentially independent from $k$, which is in sharp contrast
with the linear regime behaviour, for which the imaginary
frequency is mainly a linear function of $k$.
It can be instructive to compare the characteristic times
$\tau_{MF} = \hbar/\lambda_{\infty}$,
which defines the time scale of the divergence between mean
field trajectories, with the single particle characteristic time
$\tau_{sp}=\hbar/E_F$, being $E_F$ the
Fermi energy at the given density. This is done in table I for a
set of densities. One can see that for densities $\rho \leq 0.4\rho_0$
the divergence time is smaller than the single particle time, and
therefore in this region the notion itself of mean field ceases to
have any validity. In this region it appears more appropriate to
speak of a fully many particle chaotic behaviour \cite{zel93}.
It has to be stressed that, despite the density is small,
the mean distance between particles is only slightly larger
than at normal density,
and therefore the particles can still strongly interact.
In central collisions between heavy ions at intermediate energy
nuclear matter is expected to be compressed at the initial stage
of the reaction, until a highly excited composite nuclear system
is formed. Numerous calculations \cite{ber88}, based
on kinetic equations as well as on molecular dynamic models
\cite{aic85}, show that the momentum distribution is close to
spherical symmetry and not too far from thermodynamic equilibrium.
In this first stage of the reaction
the dynamics is insensitive to small modifications of the initial
conditions. Once the maximum compression is reached, nuclear matter
starts to expand and can merge into the spinodal region. At this
point the dynamics should change dramatically. According to the
results presented in this paper, chaotic dynamics sets in, and
the corresponding large density fluctuations dominate the cluster
formations. The degree of chaoticity and fluctuations depends of
course on the details of the dynamics which
brings the reaction inside the spinodal region. The ratio between
the expansion characteristic time and $\tau_{MF}$ could indicate
whether chaos has the time to develop.
The multifragmentation regime should appear
in the energy interval where the chaoticity and the corresponding
fluctuations are strong enough to produce several IMF.
A confirmation of this scenario could be obtained by means of
numerical simulations of heavy ion reactions.
Unfortunately, at the moment,
the extension of the calculations presented in
this paper to the case of collisions between heavy ions
appears  problematic even in two dimensions.
Thus one should find the signature of chaotic dynamics using more
phenomenological approaches.
The study of fragment size
fluctuations and their statistical properties \cite{gro93},
in particular
along the lines of the intermittency
analysis \cite{plo90}, can be one of
the viable possibilities. However, the problem is still completely
open and demands a careful analysis, both theoretically and
experimentally.

\begin {table}
\caption{Comparison between the characteristic times
$\tau_{MF}=1/\lambda_{\infty}$
for the mean field  and $\tau_{sp}= \hbar/ E_F$
for the single particle at different
values of the nuclear density. See text.}
\label {tab1}

\begin{tabular}{|ccc|}
{}~~~~~~  $ \rho/\rho_o$ ~~~ & ~$\tau_{sp}$  ~~~    & ~$\tau_{MF}$ ~~~~~\\
    ~~~~~~                 &~$fm/c$    ~~~~~~     &~$fm/c$ ~~~~~      \\
\hline
    ~~~~~~ 0.7~            & ~7.76     ~~~~ & ~1538. ~~~~~ \\
    ~~~~~~ 0.6~            & ~9.06     ~~~~ & ~28.57 ~~~~~ \\
    ~~~~~~ 0.5~            & ~10.87    ~~~~ & ~13.34 ~~~~~ \\
    ~~~~~~ 0.4~            & ~13.59    ~~~~ & ~9.92 ~~~~~  \\
    ~~~~~~ 0.3~            & ~18.12    ~~~~ & ~9.06 ~~~~~ \\
    ~~~~~~ 0.2~            & ~27.17    ~~~~ & ~10.00 ~~~~~ \\
\end{tabular}
\end{table}

\begin{figure}
\caption{ The evolution for
two close trajectories (dotted and full curve) for the
nuclear matter density profile is illustrated
as a function of time. Two different cases are
shown for  values of the initial density,
(a) $\rho/\rho_0$=1 and (b) $\rho/\rho_0$=0.5.
While in case (a) the infinitesimal
initial difference keeps constant in time and it is hardly
visible, in  case
(b) it grows exponentially.
Note the different scale used in fig.1(a) and 1(b).
}
\end{figure}

\begin{figure}
\caption{Power spectrum in momentum space of the density profile
shown in fig.1(b) as full curve. }
\end{figure}

\begin{figure}
\caption{(a)
The behaviour of $\lambda(t)$ is plotted vs. time for different
values of the initial density. The constant value obtained
for $\rho/\rho_0$=0.4, 0.5, 0.6 allows to
 define the Lyapunov exponent $\lambda_{\infty}$ and
to classify the chaoticity of
the mean  field at T=3 MeV. At variance for $\rho/\rho_0$=0.7
$\lambda(t)$ is a decreasing function of t and the mean field
has a regular linear evolution. (b)
The Lyapunov exponent $\lambda_{\infty}$ is displayed for
different values of the initial density. See text.}
\end{figure}


\begin{references}
\bibitem{gro82} D.H.E. Gross and X.Z. Zhang , Phys
Lett. {\bf 161 B }, 47 (1982);
 J.P.Bondorf, R.H.Donangelo, I.N.Mishustin, C.J.Pethick, H.Schultz and
 K.Sneppen, Nucl. Phys. {\bf A443}, 321 (1985); G.Peilert, H.St\"ocker,
 J.Randrup and W.Greiner, Phys. Lett. {\bf B260}, 271 (1991);
 C.Dorso and J.Randrup, preprint LBL--33120; M.Colonna, M.Di Toro,
 V.Latora, Nucl. Phys. {\bf 545}, 111c (1992).

\bibitem{mor92} L.Moretto, K.Tso, N.Colonna, G.Wozniak, Phys. Rev.
 Lett. {\bf 69}, 1884 (1992).

\bibitem{bau92} W.Bauer, G.Bertsch and H.Schultz, Phys. Rev. Lett.
 {\bf 69}, 1888 (1992).

\bibitem{hei93} H. Heiselberg C.J. Pethick and D.G. Ravenhall
 Phys. Rev. Lett. {\bf 61}, 818 (1988), and Ann. Phys. {\bf 223},
 37 (1993); M.Colonna, Ph.Chomaz and J.Randrup, GANIL preprint 1992.

\bibitem{ben76} G.Benettin, L.Galgani and J-M.Strelcyn,
Phys. Rev. {\bf A14}, 2338 (1976); A.J.Lichtenberg and M.A.Lieberman,
{\it Regular and Stochastic motion}, Springer--Verlag (1983).

\bibitem{bur92} G.F. Burgio, Ph. Chomaz and J. Randrup, Phys. Rev. Lett.
{\bf 69}, 885 (1992); G.F. Burgio, Ph. Chomaz and M. Colonna, Catania
preprint 1993.

\bibitem{gre87} C.Gr\'egoire, B.Remaud, F.S\'ebille, L.Vinet and
 Y.Raffray, Nucl. Phys. {\bf A465}, 317 (1987).

\bibitem{ber88} G.Bertsch and S.Gupta, Phys. Rep. {\bf 160}, 189 (1988);
A. Bonasera, F. Gulminelli and J. Molitoris, Phys. Rep. in preparation.



\bibitem{nota93}
It can be instructive to analyze what should result for $\lambda(t)$
if the linear regime would be valid, adopting the present
procedure of modifying the initial conditions. In this case
the variation $\Delta\rho$ of the density is equivalent to the addition
of a mode at $k = 0$. In the linear regime this mode has zero
frequency and propagates independently from the mode $k$, therefore at
time $t$, one has approximately
$d(t) \, \approx \, \Delta\rho \, +\, | \Delta\rho_{1} \left(\exp
(\omega + \Delta\omega)t \, -\, \exp \omega t \right) |$,
where $\omega$ is the growth rate of the mode $k$ at the density
$\rho$ and $\omega + \Delta\omega$ the corresponding one at
$\rho + \Delta\rho$.
For small enough $t$ , and not too close to the spinodal boundary,
$\lambda(t)\, =\, \log(d(t)/\Delta\rho)/t \approx
\Delta\rho_{1} \Delta\omega /\Delta\rho \approx
\omega (\Delta\rho_{1}/\rho)$, which is negligibly small, since
$\Delta\rho_{1} << \rho$. The value of $\lambda(t)$ keeps close to
zero, for $t \leq \omega^{-1}$, and then after it increases towards
the value $\omega$, which can be considered the Lyapunov exponent
of the mode $k$ in the linear regime. The actual behaviour
of $\lambda(t)$ in the spinodal zone is quite different  and this
demonstrates that the mean field dynamics cannot be described
within the linear approximation in this region.

\bibitem{zel93} V.G.Zelevinsky, Nucl. Phys. {\bf A553}, 125c (1993).

\bibitem{aic85} Aichelin and H.St\"ocker, Phys. Lett.
 {\bf B163}, 59 (1985)

\bibitem{gro93} D.H.E. Gross, Nucl. Phys. {\bf A553}, 175c (1993);

\bibitem{plo90} M.Ploszajczak and A.Tucholski, Phys. Rev. Lett.
 {\bf 65}, 1539 (1990).

\end{references}
\end{document}